\documentstyle[colordvi,prl,twocolumn,aps,epsfig,epsf]{revtex}

\begin{document}

\draft

\title{Non--Universal Behavior of the $k$--Body Embedded Gaussian
  Unitary Ensemble of Random Matrices}

\author{L. Benet, T. Rupp and H. A. Weidenm\"uller}

\address{Max--Planck--Institut f\"ur Kernphysik, D--69029
Heidelberg, Germany}

\date{\today}

\maketitle

\vspace*{-0.5cm}

\begin{abstract}

Using a novel approach, we investigate the shape of the average
spectrum and the spectral fluctuations of the $k$--body embedded
unitary ensemble in the limit of large matrix dimension. We identify
the transition point between semicircle and Gaussian shape. The
transition also affects the spectral fluctuations which deviate from
Wigner--Dyson form and become Poissonian in the limit $k\ll m \ll l$.
Here $m$ is the number of Fermions and $l$ the number of degenerate
single--particle states.
\end{abstract}


{\it Introduction}. The stochastic behavior displayed by spectra and
wave functions of quantum many--body systems (atoms, molecules, atomic
nuclei, quantum dots) is usually and successfully modelled in terms of
canonical random--matrix theory (RMT)~\cite{bro81,guh98}. However,
this type of modelling is not completely realistic: All the
above--mentioned many--body systems are effectively governed by one--
and two--body forces, while canonical RMT is tantamount to assuming
many--body forces between the constituents. Thus, a stochastic
modelling of the one-- and two--body interaction would yield a much
smaller number of independent random variables than used in RMT. For
instance, the number of independent two--body matrix elements in a
shell--model calculation in atoms or nuclei is typically much smaller
than the dimension $N$ of the matrices involved, while the number of
independent random variables in RMT is of order $N^2$. This difference
poses the question whether a more realistic stochastic modelling of
many--body systems might yield results which differ from RMT
predictions. The question was addressed in the 1970's with the help of
numerical simulations using matrices of fairly small dimensions. The
main results were: In a certain limit, the average level density does
not have the shape of a semicircle but is Gaussian; the ensembles are
neither stationary nor ergodic; unfolding of the spectra yields
Wigner--Dyson spectral fluctuation properties, see Refs.~\cite{fre71}.
Interest in model Hamiltonians with random two--body interactions has
resurged in recent years in several areas of many--body physics (see
Ref.~\cite{kot00} and further references therein), and the question
of possible further differences between such models and RMT has
resurfaced. It is the purpose of this letter to shed new light on this
question. In particular, we will show that for certain parameter
values, the spectral fluctuation properties of realistic models differ
from those of RMT. We discuss the implications of this novel and
surprising result.

Focussing attention on the case of unitary symmetry, we use a
paradigmatic model, the $k$--body embedded Gaussian random ensemble
EGUE($k$) introduced by Mon and French~\cite{mon75}. For this
ensemble, a number of analytical results has been obtained
\cite{mon75,bro81,ver84}. However, as emphasized in Ref.~\cite{bro81},
page 413, a comprehensive analytical approach to the spectral
fluctuation properties of the ensemble is still lacking. This is due
to the fact that, in contrast to RMT, embedded ensembles do not
possess the (orthogonal or unitary) invariance in Hilbert space which
is so essential for the successful analytical treatment of RMT.

We use three different methods: The supersymmetry approach, the
``binary correlation approximation'' of Mon and French~\cite{mon75},
and the construction of two ``limiting ensembles''. We also make use
of two fundamental and novel results on the second moment: The
eigenvector expansion and duality. We compare our results with those
of the Gaussian unitary ensemble of random matrices (GUE).

{\it Definitions}. Our Hilbert space is spanned by $N = {l \choose m
  }$ Slater determinants $|\mu\rangle$, $\mu = 1,\ldots,N$, obtained by
distributing $m$ spinless Fermions over $l$ degenerate single--particle
states. The ratio $f=m/l$ is the filling factor. Using standard
creation and annihilation operators $a_j^{\dagger}$ and $a_j$, $j =
  1,\ldots,l$, we have $|\mu\rangle = \prod_{s=1}^m a_{j_s}^{\dagger}
  |0\rangle$. Here $|0\rangle$ is the vacuum state. The creation
operators are written in ascending order of the indices $j_s$. The
$k$--body interaction $V_k$ has the form ($k = 1,\ldots,m$)
\begin{equation}
\label{eq1}
V_k = {\sum}^\prime v_{j_1,\ldots,j_k;i_1,\ldots,i_k}
      a_{j_1}^{\dagger} \ldots a_{j_k}^{\dagger} a_{i_k} \ldots
      a_{i_1} \ .
\end{equation}
Here and in Eqs.~(\ref{eq2}), (\ref{eq10}), the prime on the summation
sign indicates that the sums over all indices run from 1 to $l$ with
the constraint that $j_1 < \ldots < j_k$ and $i_1 < \ldots < i_k$. The
$k$--body interaction matrix $v_{j_1,\ldots,j_k; i_1, \ldots, i_k}$ is
complex Hermitean. The independent matrix elements are uncorrelated
Gaussian random variables with zero mean and a common second moment
$v^2$. The value of $\sqrt{v^2}$ determines the overall energy scale
and is set equal to unity without loss of generality. By taking matrix
elements $\langle \nu | V_k | \mu \rangle$, we ``embed'' the random
$k$--body interaction into an $m$--Fermion system. This defines
EGUE($k$). The ensemble is invariant under unitary transformations of
the single--particle states $1,\ldots,l$ and, for $k=m$, reduces to
the standard GUE. As always in RMT, we are interested in the limit
$N\to\infty$ ($l \to \infty$).

{\it Second moment}. The Gaussian distribution of the matrix elements
in Eq.~(\ref{eq1}) implies that the $\langle \nu | V_k | \mu \rangle$'s
are also Gaussian distributed random variables with zero mean value.
All spectral properties of EGUE($k$) are, therefore, determined by the
second moment
\begin{eqnarray}
\label{eq2}
A_{\mu \nu , \rho \sigma}^{(k)} &=& \overline{\langle \mu | V_k | \sigma
  \rangle \langle \rho | V_k | \nu \rangle} \nonumber \\
&=& {\sum }^\prime \langle \mu |
   a_{j_1}^{\dagger} \ldots a_{j_k}^{\dagger} a_{i_k} \ldots a_{i_1} |
   \sigma \rangle \nonumber \\
&&\quad \times \langle \rho | a_{i_1}^{\dagger} \ldots
   a_{i_k}^{\dagger} a_{j_k} \ldots a_{j_1} | \nu \rangle \ .
\end{eqnarray}
The bar denotes the ensemble average. We have $A_{\mu \nu, \rho
  \sigma}^{(k)} = A_{\rho \sigma, \mu \nu }^{(k)}$ and $(A_{\mu \nu,
  \rho \sigma}^{(k)})^{*} = A_{\sigma \rho, \nu \mu}^{(k)} = A_{\mu
  \nu, \rho \sigma}^{(k)}$. The matrix $A^{(k)}$ is Hermitean in the
pairs of indices $(\mu,\nu)$ and $(\rho,\sigma)$. Moreover, it is easy
to prove the ``duality'' relation $A^{(k)}_{\mu \nu, \rho \sigma} =
  A^{(m-k)}_{\mu \sigma, \rho \nu}$ which connects the second moments
of the $k$--body and the $(m-k)$--body interaction.

{\it Eigenvector expansion}. We construct the eigenvectors $C^{(s a)}$
and eigenvalues $\Lambda^{(s)}(k)$ of the Hermitean matrix $A^{(k)}$
satisfying $\sum_{\rho \sigma} A_{\mu \nu, \rho \sigma}^{(k)} C_{\sigma
  \rho}^{(s a)} = \Lambda^{(s)}(k) C_{\mu \nu}^{(s a)}$. Here $s=0,
\ldots, m$ and $a$ labels the degenerate eigenvectors. We are guided
by the example of the GUE where the second moment of the Hamiltonian
$H$ reads $\overline{H_{\mu \sigma} H_{\rho \nu}} = (\lambda^2/N)
\delta_{\mu \nu} \delta_{\rho \sigma}$. The two Kronecker symbols
display the unitary invariance of the GUE. The matrix $\delta_{\sigma
  \rho}$ is eigenfunction of the second moment with eigenvalue
$\lambda^2$. All traceless unitary matrices are likewise
eigenfunctions but belong to eigenvalue zero. In the present case, we
use the ansatz $C_{\mu \nu}^{(s a)} = \langle \mu | a_{j_1}^{\dagger}
\ldots a_{j_s}^{\dagger} a_{i_s} \ldots a_{i_1} | \nu \rangle $. The
label $a$ enumerates all possible distinct choices of the indices $j_1
< \ldots < j_s; i_1 < \ldots < i_s$. It is easy to check that $C^{(s
  a)}$ is eigenvector of the matrix $A^{(k)}$ if no two indices
$(j_r,i_{r^\prime})$ are equal. The corresponding eigenvalue is
\begin{equation}
\label{eq3}
\Lambda^{(s)}(k) = {m-s \choose k} {l-m+k-s \choose k} \ .
\end{equation}
When at least two indices $(j_r,i_{r'})$ are equal, $C_{\mu \nu}^{(s
  a)}$ is not an eigenfunction of $A^{(k)}$ but a linear combination
of eigenfunctions with labels $s' \leq s$. This is because an
eigenfunction $C_{\mu \nu}^{(s' a)}$ with eigenvalue $\Lambda^{(s')}(k)$,
$s' < s$, looks like a member of class $s$ when the operator defining
$C^{(s' a)}$ is multiplied by $(\sum_p a_p^{\dagger} a_p)^{s - s'} =
  m^{s-s'}$. To remove the components belonging to lower $s'$--values,
we orthogonalize (in the sense of the trace) all matrices $C_{\mu
  \nu}^{(s a)}$ in which at least two indices $(j_r,i_{r'})$ are
equal, to all matrices generated from classes $s' < s$ in the manner
just described. The resulting matrices are eigenvectors with eigenvalue
$\Lambda^{(s)}(k)$. We choose Hermitean linear combinations of the degenerate
eigenvectors which obey the orthonormality condition $\sum_{\mu \nu}
  C_{\mu \nu}^{(s a)} C_{\nu \mu}^{(t b)} = N \delta_{s t} \delta_{a
  b}$. The number $D^{(s)}$ of linear independent eigenvectors in class
$s$ is given by $D^{(0)} = 1$ and
\begin{equation}
\label{eq4}
D^{(s)} = { l \choose s}^2 - { l \choose s-1}^2 \ ; \ s \geq 1 \ .
\end{equation}
We have $\sum_{s=0}^m D^{(s)} = N^2$ showing that the eigenvectors
form a complete orthonormal set. Hence, the matrix $A^{(k)}$ possesses
an eigenvalue decomposition of the form
\begin{equation}
\label{eq5}
A_{\mu \nu, \rho \sigma}^{(k)} = {1 \over N} \sum_{s = 0}^m
\Lambda^{(s)}(k) \sum_a C^{(s a)}_{\mu \nu} C^{(s a)}_{\rho \sigma} \ .
\end{equation}
We note that the eigenvectors do not depend on the rank $k$ of the
interaction, only the eigenvalues do. Eq.~(\ref{eq3}) shows that the
sum over $s$ actually terminates at $s = m - k$. For $k = m$ only $s =
0$ contributes, and the result reduces to the GUE expression with
$\lambda^2 = \Lambda^{(0)}(m)$. Conversely, Eq.~(\ref{eq5}) extends
the GUE result to EGUE($k$) and constitutes a central result of this
paper.

{\it Moments of} $V_k$. Using the eigenvector decomposition of
$A^{(k)}$, duality, and the orthonormality of the $C^{(s a)}$'s, we
calculate the low moments of $V_k$ and the kurtosis $\kappa$ for which
we write $\kappa = 2 + Q(k,m,l)$. We recall that $\kappa = 2 (3)$ for
the semicircle (Gaussian, respectively). We find $Q(k,m,l) = (1/N)
\sum_{s=0}^{{\rm min}(m-k,k)} [(\Lambda^{(s)}(k) \Lambda^{(s)}(m-k)) /
(\Lambda^{(0)}(k))^2)] D^{(s)}.$ For $l \to \infty$ and keeping both
$k$ and $m$ fixed, we have $Q(k,m,l) \to 0$ if $2k > m$ while
$Q(k,m,l) \to {m-k \choose k}/{m \choose k}$ for $2k \leq m$. This
shows that the transition of the average spectrum from semicircle to
Gaussian shape begins (with decreasing $k$) at $2k = m$. We ascribe
the special role of the point $2k = m$ to duality. Likewise we can
show that the relative fluctuations of the first and second moments of
$V_k$ and, thus, non--ergodic features vanish for $l \to \infty$.

{\it Supersymmetry}. We calculate spectrum shape and spectral
fluctuation properties of EGUE($k$) using the supersymmetry
approach~\cite{efe83,ver85}. After averaging over the ensemble, the
integrand of the generating functional contains an exponential whose
argument depends linearly on the matrix $A^{(k)}$. With the
eigenvalue decomposition~(\ref{eq5}), the argument of the exponential
becomes a sum of squares of bilinear forms in the integration variables.
This allows us to perform the Hubbard--Stratonovich transformation.
For each pair $(s,a)$ we introduce a supermatrix $\sigma^{(sa)}$ of
composite variables. The resulting integral over the composite
variables contains the factor $\exp(-{\cal L}_{\rm eff})$. The
effective Lagrangean ${\cal L}_{\rm eff}$ is given by
\begin{eqnarray}
&& \frac{N}{2} \sum_{s a}{\rm trg} \bigl[ \sigma^{(s a)} \bigr]^2 +
  {\rm tr}_{\mu} {\rm trg} \ln \biggl[ (E - \frac{1}{2}{\epsilon} L)
  \delta_{\mu \nu} \nonumber \\
&&\qquad\qquad - \sum_{s a } \lambda^{(s)}(k) \sigma^{(s a)}
  C^{(s a)}_{\mu \nu} - J_{\mu \nu} \bigg] \ .
\label{eq7}
\end{eqnarray}
Here, $\lambda^{(s)}(k)=+\bigl[ \Lambda^{(s)}(k) \bigr]^{1/2}$. The
energy arguments $E_1$ and $E_2$ of the advanced and the retarded
Green function define $E = (1/2)(E_1 + E_2)$ and $\epsilon = E_2 - E_1$,
while $J$ stands for the source terms. The diagonal supermatrix $L$
distinguishes the retarded and advanced cases and is defined in
Ref.~\cite{ver85}. The saddle--point equation has the solution
$\sigma^{(s a)}_{\rm s.p.} = \delta_{s 0} \tau^{(0)}$ where
$\tau^{(0)}$ is the standard GUE saddle--point solution. Thus, the
saddle--point condition yields a semicircular spectrum and universal
GUE spectral fluctuations. To determine the range of validity of this
solution, we have calculated the first non--vanishing term in the loop
expansion. The expansion is obtained by writing $\sigma^{(s a)} =
\sigma^{(s a)}_{\rm s.p.} + \delta \sigma^{(s a)}$ and expanding in
powers of $\delta \sigma^{(s a)}$. We recall that for the GUE, each
term of the loop expansion vanishes in the limit $N \to \infty$ with
an inverse power of $N$. For the one--point function, we find that the
loop correction is proportional to $Q(k,m,l)$. This is consistent with
the result of the previous paragraph and reaffirms our conclusion that
the transition from semicircular to Gaussian shape sets in at $2k = m$.
It would be desirable to show that all higher terms of the loop
expansion vanish likewise asymptotically for $2k > m$ but this proof
is beyond our means. For the two--point function, the loop correction
yields non--universal spectral fluctuations of the type first
considered by Kravtsov and Mirlin~\cite{kra94}. The amplitude of this
correction (given in units of the inverse mean level spacing)
vanishes, however, for $l \to \infty$. It does so like $N^{-2}$ for $k
= m$ but only like $(\ln N)^{-2k}$ for both $k$ and $f$ fixed. We see
that for $l \to \infty$, the supersymmetry approach does not yield a
limit on the range of validity of the universal GUE spectral
fluctuations of EGUE($k$).

{\it The case} $k \ll m \ll l$. For this case where the supersymmetry
method does not yield relevant information on spectral fluctuations,
we use a modification of the ``binary correlation
approximation''~\cite{mon75,bro81}. In the average two--point function
$\overline{g(z_1)g(z_2)}$, we expand~\cite{ver84} both traced Green
functions $g$ in powers of $V_k$. We collect the terms containing
equal powers of $V_k$. The ensemble average is taken by
Wick--contracting pairs of $V_k$'s in all possible ways. We evaluate
all pairs located on the same Green function as in Ref.\cite{mon75}.
For the rest, we use simple counting arguments. We show that for $k
\ll m \ll l$, terms where $s$ pairs of $V_k$'s are not on the same
Green function, are smaller by at least a factor $l^{-sk}$ than the
terms with $s=0$. Thus, the connected part of
$\overline{g(z_1)g(z_2)}$ vanishes asymptotically compared to the
disconnected part, and the two--point correlation function
\begin{equation}
R_2(z_1,z_2) =
\frac{\overline{g(z_1)g(z_2)}}{\overline{g(z_1)}\cdot\overline{g(z_2)}}
- 1 
\end{equation}
approaches zero in the limit $k \ll m \ll l$. For similar reasons, all
higher correlation functions also vanish in the same limit, and the
spectral fluctuations become Poissonian.

The case $k=1$ with $k \ll m \ll l$ illustrates this result. Taking
first $m = k = 1$, we construct the eigenvalues $\epsilon_{\alpha}$
and eigenfunctions $\Psi_{\alpha}$ of a given realization of EGUE(1)
by diagonalization.  The $\epsilon_{\alpha}$'s obey GUE spectral
statistics, the average spectrum has semicircle shape and the
coefficients $U_{\alpha j}$ in the expansion $\Psi_{\alpha} = \sum_{j
  = 1}^l U_{\alpha j} a_j^{\dagger} | 0 \rangle$ in terms of the
single--particle basis are Gaussian distributed random variables. For
$m > 1$ and every realization of the ensemble, the eigenfunctions are
Slater determinants $\chi$ of the $\Psi_{\alpha}$'s, the eigenvalues
are sums of the $\epsilon_{\alpha}$'s, and the average spectrum is an
$m$--fold convolution of the semicircle. This shows immediately that
for $m \gg 1$, the spectrum has Gaussian shape, and that the spectral
fluctuations are Poissonian. We have also shown that the states $\chi$
are localized.

{\it Regular graphs}. Further insight into the spectral properties of
EGUE($k$) is gained by using yet another approach involving regular
graphs and limiting ensembles. A graphical representation of EGUE($k$)
is obtained by assigning to each Hilbert--space vector $| \mu \rangle$
a vertex $\mu$, and to each non--diagonal element $\langle \nu | V_k |
\mu \rangle$ which does not vanish identically, a link connecting the
vertices $\nu$ and $\mu$. The diagonal matrix elements $\langle \mu |
V_k | \mu \rangle$ are represented by loops attached to the vertices
$\mu$. The number of vertices is $N$. The number $M$ of links
emanating from a given vertex is the same for all vertices and given
by
\begin{equation}
\label{eq8}
M = \sum_{s = 1}^k {m \choose s} {l-m \choose s} \ .
\end{equation}
For $k < m$, we have $M < N - 1$ while $M = N - 1$ for $k = m$. The
resulting graphical structure is called a ``regular graph'' in the
mathematical literature. The total number $P$ of links is given by $P
= (1/2) M N$. The number $K$ of uncorrelated matrix elements of
EGUE($k$) is given by $K = {l \choose k}^2$. It is interesting to
study the ratio $K/P$. For fixed $m$ and $f \leq 1/2$, $K/P$ grows
monotonically with $k$, starting out with very small values and
reaching the limit $2 N / (N - 1)$ for $k = m$. This suggests that
deviations of EGUE($k$) from universal GUE behavior are caused by the
fact that the number of independent random variables is too small in
comparison with the number of links, making it impossible for the
system to become thoroughly mixed.

{\it Limiting ensembles}. To test this hypothesis, we have constructed
and analyzed two limiting ensembles. The first, EGUE$_{\rm min}(k)$,
is given by the matrix elements $\langle \nu | V_k^{\rm min} | \mu
\rangle$ of the interaction $V_k^{\rm min} = v {\sum}^{\prime}
a_{j_1}^{\dagger} \ldots a_{j_k}^{\dagger} a_{i_k} \ldots a_{i_1}$.
The factor $v$ is a Gaussian complex random variable. The ensemble
EGUE$_{\rm min}(k)$ has the same graphical representation as EGUE($k$)
but possesses only a single random variable. We succeeded to prove that
for $k=1$ and $k=m$, EGUE$_{\rm min}(k)$ is fully integrable,
has a Gaussian average spectrum and spectral fluctuations which are
not of GUE type. We are convinced that these properties hold for all $k$.
The second limiting ensemble, EGUE$_{\rm max}(k$), carries the maximum
number of uncorrelated random variables consistent with the graph
structure of EGUE($k$) and has Hilbert--space matrix elements given by
\begin{equation}
\label{eq10}
v_{\nu \mu} {\sum}^{\prime} \langle \nu |
  a_{j_1}^{\dagger} \ldots a_{j_k}^{\dagger} a_{i_k} \ldots a_{i_1} |
  \mu \rangle \ .
\end{equation}
The matrix $v_{\nu \mu}$ is complex Hermitean. Elements not connected
by symmetry are uncorrelated complex Gaussian random variables with
mean value zero and variance $\overline{v_{\nu \mu} v_{\nu' \mu'}} =
\delta_{\mu \nu'} \delta_{\nu \mu'}$. Using supersymmetry, we have
shown that EGUE$_{\rm max}(k$) has an average spectrum of semicircle
shape, that the spectral fluctuations are of universal GUE type, and
that for $l \to \infty$ the leading term of the loop expansion
vanishes both for the one--point and for the two--point functions. We
conclude that the spectral properties of EGUE$_{\rm max}(k$) coincide
with those of GUE.

{\it Conclusions}. We have studied the shape of the average spectrum
and the eigenvalue fluctuations of the embedded ensemble EGUE($k$) in
the limit of infinite matrix dimension, attained by letting the number
$l$ of degenerate single--particle states go to infinity. We have shown
that for sufficiently high rank $k$ of the random interaction ($2k >
m$ where $m$ is the number of Fermions), EGUE($k$) behaves generically:
The spectrum has semicircle shape, and the eigenvalue fluctuations
obey Wigner--Dyson statistics. This does not come as a surprise. A
transition to a different regime takes place at or near $2k = m$. It
has long been known that the average spectrum changes into Gaussian
shape, although the point of departure from the semicircle shape was
not known previously. We have presented conclusive evidence that in
addition --- and contrary to general expectations --- the level
fluctuations also change and are not of Wigner--Dyson type for $2k
\lesssim m$. In the extreme case $k \ll m \ll l$, the spectral
fluctuations are Poissonian, and the eigenfunctions are likely to
display localization in Fock space. We cannot pin down precisely the
$k$--value where such change occurs nor can we penetrate deeply into
the intermediate regime. This is not surprising as we do not know of
any other case where such an aim would have been achieved. We cannot
even say definitively whether the transition from Wigner--Dyson to
Poissonian statistics is smooth or sudden. But we have circumstantial
evidence for a smooth transition: (i) The non--universal fluctuations
calculated from the loop correction set in smoothly. (ii) In the case
$k=1$, the transition from GUE (for $m=1$) to Poisson behavior (for $k
\ll m$) is smooth. (iii) The ratio $K/P$ of the number of uncorrelated
random variables over the number of links changes smoothly with $k$
for fixed $m$.

Formally and using diagrammatic language, we ascribe the deviations
from universal GUE behavior to the fact that with decreasing $k$,
intersecting Wick contraction lines gain increased weight. Universal
GUE results are obtained whenever such contributions are negligible.
This apparently is the case for $2k > m$. We ascribe the special role
of the transition point $2k = m$ to duality.

Physically, our results can be understood in terms of the ratio $K/P$
of the number of uncorrelated random variables over the number of
links.  We have shown that if all links were to carry uncorrelated
random variables, the ensemble would have GUE spectral fluctuations.
Conversely, if all links were to carry the same random variable, the
ensemble would be completely integrable and display Poissonian
statistics. These statements hold for all values of $k$. The actual
situation is located between these two limits. EGUE(1) is closest to
the integrable case, and EGUE($m$) corresponds to the GUE. This shows
that deviations from GUE behavior are not caused by the number of
zeros in the matrix representation of the interaction but are strictly
due to correlations between the matrix elements carried by the links.

As mentioned in the introduction, numerical simulations have shown
good agreement between the spectral fluctuations of EGUE($k$) and
those of GUE. We ascribe this result to the fact that the dimensions
of the matrices used were quite small. It is easy to see that the
ratio $K/P$ is relatively close to unity for typical values like
$k = 2$, $m = 8$, and $l = 20$. It is only in the limit $l \gg 1$ that
$K/P$ becomes very small, resulting in significant deviations of
EGUE($k$) spectral fluctuation behavior from Wigner--Dyson statistics.

Details, further results and the extension to EGOE($k$), the orthogonal
case, are given in Ref.~\cite{ben00}.

{\it Acknowledgment}. We are grateful to O. Bohigas and T.H. Seligman.
Both helped us with many questions and suggestions and were actively
involved in some aspects of this work.

\end{document}